\documentclass{article}
\usepackage[utf8]{inputenc}
\usepackage{amsmath}
\usepackage{graphicx}
\usepackage{geometry} 
\usepackage[
    backend=biber,
    style=numeric,
  ]{biblatex}
\usepackage{lineno}

\title{The Point of No Return: \\
\large Evolution of Excess Mutation Rate is Possible Even for Simple Mutation Models}
\author{Brian Mintz and Feng Fu\\
Department of Mathematics, Dartmouth College, Hanover, NH 03755, USA} 
\date{April 2022}

\begin{document}
\maketitle

\begin{abstract}
    Under constant selection, each trait has a fixed fitness, and small mutation rates allow populations to efficiently exploit the optimal trait. Therefore it is reasonable to expect mutation rates will evolve downwards. However, we find this need not be the case, examining several models of mutation. While upwards evolution of mutation rate has been found with frequency or time dependent fitness, we demonstrate its possibility in a much simpler context. This work uses adaptive dynamics to study the evolution of mutation rate, and the replicator-mutator equation to model trait evolution. Our approach differs from previous studies by considering a wide variety of methods to represent mutation. We use a finite string approach inspired by genetics, as well as a model of local mutation on a discretization of the unit intervals, handling mutation beyond the endpoints in three ways. The main contribution of this work is a demonstration that the evolution of mutation rate can be significantly more complicated than what is usually expected in relatively simple models.
    
    \vspace{0.5cm}
    
    \textbf{Keywords:} Adaptive dynamics, replicator-mutator equation, mutation rate evolution. 
\end{abstract}

\section{Introduction}


Evolution occurs in a population through a repeated process of mutation, which introduces new traits, and selection, where traits leading to a higher reproduction rate outcompete less fit traits. Often, mutation is thought of as the result of defects in the reproduction process. This can take place in a genetic context, for example errors in DNA or RNA transcription, or a cultural one, such as imperfect language acquisition or offspring behaving differently than their parents. Crucially, the mutation rate is itself subject to evolution because of the existence of error-correcting genes, in the genetic context, and the degree to which a society will change its belief and practices is itself a part of culture. As well as acting between generations, mutation can be thought of within a generation as, albeit random, exploration, for example by varying hunting strategies or organization structures. The other mechanism, selection, can be constant or change in some deterministic manner. For example, the fitness of a trait may depend on the proportion of other traits in a population, that is, selection is frequency dependent (\cite{freq_1}, \cite{freq_2}, \cite{freq_3} and \cite{freq_4}). Alternatively, selection may be time dependent; perhaps a trait was initially viable but now is not (\cite{time_1}, \cite{time_2}, \cite{time_3}, \cite{time_4}, and \cite{time_5}). This wide variety of contexts can all be encoded in the mathematical framework of evolutionary dynamics. By analyzing this, we can discover interesting effects and provide insights into the paths taken by evolution. 

One of the most basic models of evolution considers a finite number of traits, each with a constant fitness $f_i$, effectively their rate of reproduction. A dependence on time or population composition can be easily incorporated by making the fitness of trait $i$ a function of these, but doing so greatly complicates the subsequent analysis. It is also helpful to assume a infinitely large, well mixed population, as this allows the dynamics to be described by a system of ODE's, the \textit{replicator equation} $\frac{d}{dt} x_i = x_i (f_i-\phi)$ for each trait $i$ with proportion $x_i$ in the population, and $\phi \equiv \sum_i x_i f_i$ the average fitness (introduced by the normalization $\sum_i x_i = 1$). Under this model, a population adapts to a fitness landscape, with the most fit trait becoming dominant (\cite{eqns_of_life}, \cite{RD_1}, \cite{RD_2}, \cite{RD_3}, and \cite{RD_4}). This equation encodes selection, and to account for mutation we can introduce a matrix $Q(u)$, whose $ij^\textrm{th}$ entry is the transition probability from trait $j$ to trait $i$ given a parameter $u$ for mutation rate (while $Q(u)$ technically gives the degree, not rate, of mutation, these are interchangeable given a unit time step). Now the dynamics are given by the \textit{replicator-mutation equation} \begin{equation}\label{RME}
    \frac{d}{dt} \vec{x} = (Q(u)F-\phi I)\vec{x} 
\end{equation} where $\vec{x} = (x_1, x_2, ..., x_n)$, $F$ has $ii^\textrm{th}$ entry the fitness of trait $i$, and $I$ is the $n \times n$ identity matrix (\cite{RME_1}, \cite{RME_2}, \cite{RME_3}, \cite{RME_4}, \cite{RME_5}, \cite{RME_6}, \cite{RME_7}, and \cite{RME_8}). While this forms a complete description of the evolution of traits under a constant mutation rate, it does not specify how the mutation rate itself evolves. To do this, it helps to assume a separation of timescales between the evolution of a trait and the mutation rate \cite{main_ref}. This makes it sufficient to analyze how an invading mutation rate will perform in a population at equilibrium, formalized by \textit{adaptive dynamics}.

Adaptive dynamics is a technique to predict evolutionary outcomes under small mutations (\cite{AD_1}, \cite{AD_2}, \cite{AD_3}, \cite{AD_4}, \cite{AD_5}, \cite{AD_6}, \cite{AD_7}, and \cite{AD_8}). In this framework, mutants appear infrequently enough that competition between invaders and residents, who have reached equilibrium, occurs before subsequent mutants arise. The likelihood of fixation of invading mutants with one-dimensional trait $y$ into a homogeneous population with trait $x$ is given by the \textit{invasion fitness} $f_x(y)$, whose exact form depends on the particular application. Since mutations are small enough, the linear approximation $f_x(y) \approx f_x(x)+(y-x)\frac{\partial}{\partial y} f_x(y)\rvert_{y=x}$ is accurate. Since $f_x(x) = 0$, this means invasion can only occur if $y-x$ is the same sign as $\frac{\partial}{\partial y} f_x(y)\rvert_{y=x}$. Thus, up to a constant, the rate of change of the trait is given by \begin{equation}\label{AD}
    \frac{d}{dt} x = D(x) := \frac{\partial}{\partial y} f_x(y)\biggr\rvert_{y=x}
\end{equation} This will have equilibria when $D(x)=0$, with stability determined by the sign of $\frac{d}{dx}D(x)$.

Combining these two techniques, \cite{main_ref} determines that the appropriate invasion fitness of a small invading population with mutation rate $u'$ and distribution $x'$ into a resident population with mutation rate $u$ and distribution $\tilde{x}$ is \begin{equation}\label{IF}
    f_{u,\tilde{x}}(u',x') = \lambda_{\max}(u',\tilde{x}) - \phi(\tilde{x})
\end{equation} where $\lambda_{\max}(u',\tilde{x})$ is the maximum eigenvalue of $Q(u')F(\tilde{x})$, and $\phi(\tilde{x})$ is the average fitness of the distribution $\tilde{x}$, and $\tilde{x}$ is an equilibrium distribution of equation \eqref{RME}. It is important to note that a positive invasion fitness does not necessarily imply the invaders will replace the resident population, though this is a likely proposition.

With this mathematical framework, we are able to ask some profound questions about the evolution of mutation rate. What is the best mutation rate in a given setting? Can a population evolve towards a non-optimal, but stable, mutation rate? What are the effects of various design choices on the outcome of the system? Previous work has shown interesting results for time and frequency dependent fitness, but is this necessary for such behavior? Intuitively, one might expect low mutation rates to be preferred, as they cause a population to deviate less from an optimum. However, they can also lead to a population being stuck at a local, but not global, maximum. This work investigates these questions, most notably finding instances of nontrivial mutation rate evolution in several plausible models of mutation. Surprisingly, we demonstrate that these effects can occur even in the simple context of constant selection. 


\section{Methods and Models}
\begin{figure}[h]
\centering
 \includegraphics[width=6.5in]{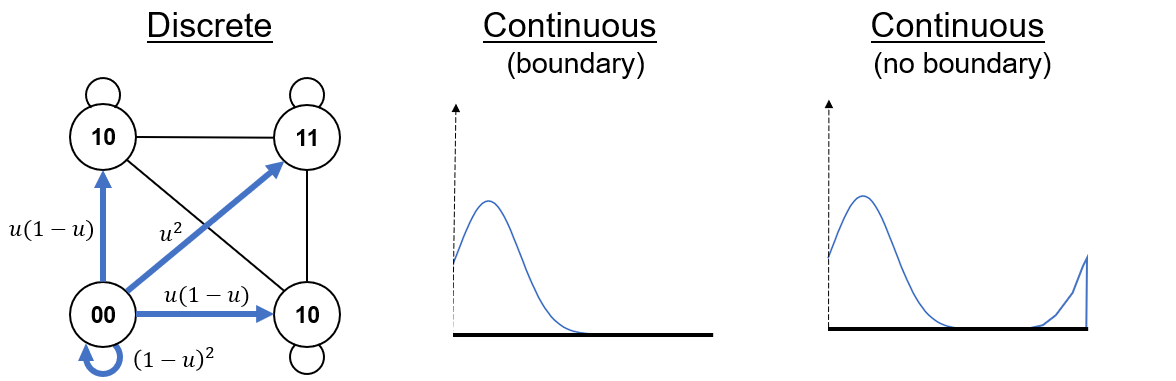}
\caption{Diagrams of three models for mutation. First is finite strings on a finite alphabet with independent, uniform mutation on each letter. The next two show local mutation on continuous traits. Mutations outside the boundary is truncated in the middle plot, and wraps around in the final plot. This removes the boundaries in the last model, making the trait space a circle. The local mutation is shown as a normal distribution, but it could be any curve, for example, an indicator function of width $u$.}
\label{domains}
\end{figure}

Whereas most models assume traits are unrelated, this work uses various notions of locality to model mutation, which should be more common to 'close' than 'far' traits. We represent mutation in three ways, depicted in Fig.~\ref{domains}. 

One option is motivated by genetics, considering traits a collection of genes, each with some finite number of alleles that mutate independently. For simplicity, one may take all genes to have the same number of alleles, and mutation to be symmetric (\cite{finite_strings_1} and \cite{finite_strings_2}). Then this model becomes finite strings on a finite alphabet, with letters independently changing with some probability $u$ to one of the other letters uniformly at random. This is depicted in the first part of Fig.~\ref{domains}, in the case of binary strings of length two. Here we see the relative mutation probabilities from a given trait. One benefit of this approach is the closed form entries of $Q(u)$, namely entry $ij$ is $(u/(k-1))^d(1-u)^{n-d}$ where $n$ is the length of the string, $k$ is the size of the alphabet, and $d$ is the number of positions that have different values in the strings $i$ and $j$. 

Another approach is to consider traits as representing some bounded one dimensional quantity, such as an organism's weight or the length of an appendage. This could also be any behaviour that can sensibly be assigned a number, such as a preference for cooperation. In this setting, it is natural to represent mutation as a local process, since one should expect mutants not to differ too much from their parents. We formalize this idea by using a spread kernel that determines how offspring deviate from their parents. For example, this might be the normal distribution centered around the parents trait with standard deviation $u$, or a uniform probability to any trait within $u/2$ of the parent's trait. This is shown in the middle of Fig.~\ref{domains}, where the curve represents the trait distribution of offspring of a given individual. Since the theoretical framework is based on a finite collection of traits, it is necessary for the trait space to be bounded. Consequently, mutation outside of these bounds must be handled in some way. Two reasonable approaches are to either truncate out of bound values, or accumulate them and add the sum to the lowest possible trait. These often result in similar effects, so truncation is considered for simplicity, though the results for accumulation are also given. This is displayed in Fig.~\ref{domains}, where the portion of the curve outside the interval is omitted. Alternatively, we also consider making the boundary periodic, resulting in a circular trait space. This is depicted in the last part of Fig.~\ref{domains}, note how the portion of the normal curve beyond the boundary is translated through the other endpoint. 

The model used will get encoded in the matrix $Q(u)$. The results of \cite{main_ref} can then be applied once one specifies the fitness function. In general this is quite difficult, mainly due to determining the limiting distribution $\tilde{x}$. At equilibrium, the right hand side of equation \eqref{RME} will be zero, yielding $Q(u)F(\tilde{x},t) = \phi \tilde{x}$, that is, $\tilde{x}$ is an eigenvector of $Q(u)F(\tilde{x},t)$. Since the entries of $F$ themselves depend on the unknown $\tilde{x}$, and a time variable $t$, this is challenging to solve, so it is more effective to numerically solve the system of differential equations given by equation \eqref{RME}, for example by the forward Euler method. However, if fitness does not depend on the trait distribution or time, as in constant selection, then $F(\vec{x}, t)$ is simply a constant matrix $F$, making $\tilde{x}$ the the eigenvector corresponding to the maximum eigenvalue of $Q(u)F$. This may be solved explicitly, yielding a faster and more accurate solution for $\tilde{x}$. Further, this eigenvalue is the growth of the entire population, which is the same as the average rate of reproduction $\phi$. Thus the invading mutation rate $u'$ will have positive invasion fitness, for a resident $u$, when $\lambda_{\max}(Q(u')F) > \lambda_{\max}(Q(u)F)$. That is, the mutation rate will evolve in the direction of higher values of $\lambda_{\max}(Q(u)F)$, so local maxima/minima of this curve will be stable/unstable equilibria. The matrix $Q(u)F$ and its eigenvalues are calculated using Matlab, the code for which is available Github repository https://github.com/bmDart/Evolution-of-Mutation-Rate. 



\section{Results}

For each model of mutation considered, we identify conditions leading to regions where increases in mutation rate are favored, though the optimal mutation rates were those closest to zero.

\begin{figure}[h]
\centering
 \includegraphics[width=6.5in]{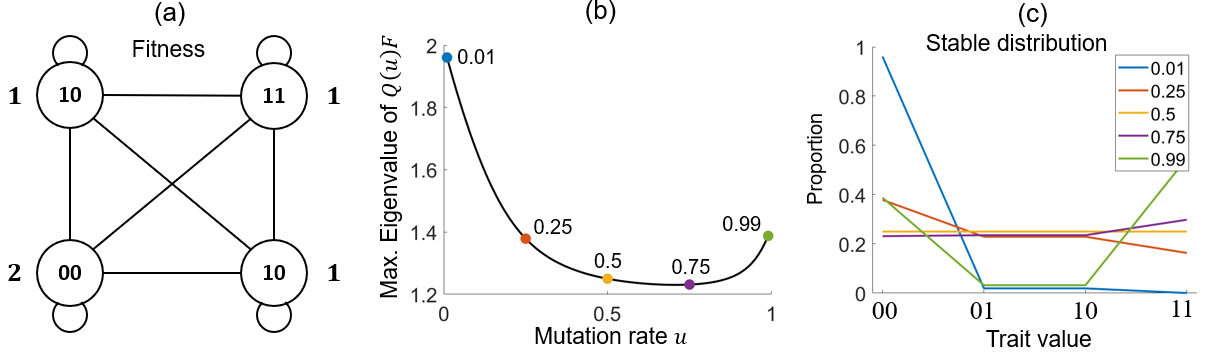}
\caption{For the given fitness function (a), we see mutation is eventually favored, the maximum eigenvalues increase with mutation rate (b). The final panel shows that the stable distributions flatten with increasing mutation, then become bimodal. }
\label{discrete}
\end{figure}

In the simplest nontrivial case of finite strings on a finite alphabet, binary strings of length two, one often sees a non-monotonic curve. This was observed in \cite{finite_strings_1} using binary strings of length three and a related fitness function. The authors explain that the large fitness for mutation rates near one result from an approximate cycle between states with mostly ones and states with mostly zeros. Therefore, if those states are fitter, higher mutation rates are selected for. A similar effect is seen for longer strings and larger alphabets, but is less pronounced. This is likely because the mutation is less likely to create such cycles. Interestingly, this effect disappeared if mutation is instead allowed to result in any letter, including the starting letter (that is, when there is a $(u/k)^d(1-u+u/k)^{n-d}$ probability of transitioning between strings with $d$ different characters, where $n$ is the length of the string and $k$ is the size of the alphabet). Fig.~\ref{discrete} shows that for a certain fitness, mutation rates above 0.5 are favored. Specifically, the fitness highest at the trait 00 and uniform elsewhere, panel (a). The maximum eigenvalue of $Q(u)F$ is plotted in (b), where one sees an increase near one, leading to upward evolution of mutation rate. One sees the local mimimum occurs around three quarters, so higher mutation rates will only evolve if the initial value is significantly large, reflecting a highly inaccurate replication process. Since the goal of organisms is to replicate themselves, usually with an accuracy far better than half, such a scenario is unlikely despite being mathematically possible. Lastly, we see the limiting distributions in panel (c), which confirm the explanation given above; for large mutation rates the population is balanced between the fittest trait 00, and its opposite 11. 


\begin{figure}[h]
\centering
 \includegraphics[width=6.5in]{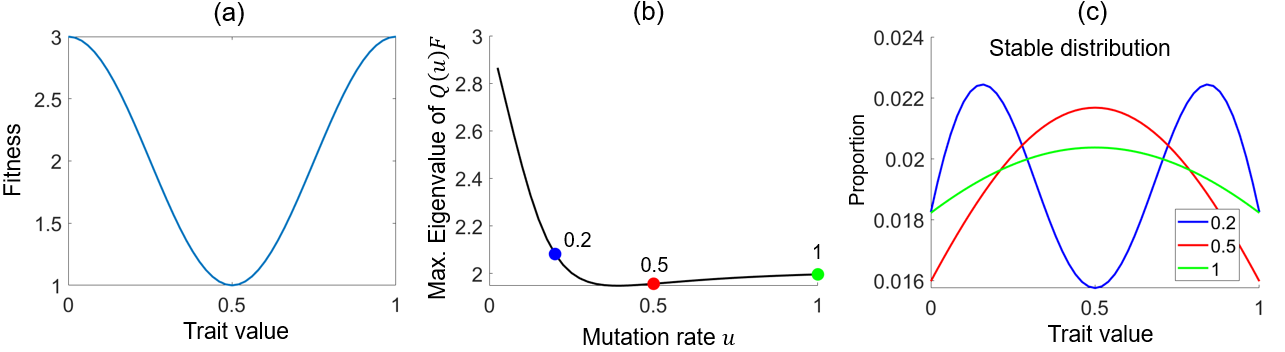}
\caption{When fitness is maximal near the boundaries, (a), accumulation near the endpoints can lead to increase fitness with mutation rate, seen by the increase in (b). In (c), we see that for intermediate mutation rates, the distribution has a peak in the center. This decreases with larger mutation rate, increasing overall fitness. These results are derived in the truncation model, though this effect is more pronounced when accumulation is used. }
\label{trunc}
\end{figure}

Next, when out of bound mutants are accumulated, the endpoints contain most of the population for large mutation rates. Therefore if the endpoints have high fitness, it makes sense that large mutation rates can be selected for. More surprising is that this still holds when mutation is truncated at the boundary, though to a lesser extent. This is unexpected, as mutation near the boundaries is more towards the middle, so there will not be the same clustering at the endpoints. However, since traits in the center are mutated to more often than traits at the boundary, the distribution will be peaked in the center. This effect diminishes with larger mutation rate, so if traits at the boundary are more fit, mutation rate can increase. An example of this is given in Fig.~\ref{trunc}. Like Fig.~\ref{discrete}, three panels show the fitness function, maximum eigenvalue as a function of $u$, and stable distributions. Here, we see a local minimum, and therefore an evolutionary unstable, mutation rate around 0.4. As explained, panel (c) shows that increasing the mutation rate from 0.5 to 1 flattens the distribution, causing more of the population to be at the optimum trait zero or one. A similar curve can be observed even in a very coarse discretization. Indeed one only needs to use the points 0,  0.5, and 1, allowing one to explicitly solve for the maximum eigenvalue as a function of $u$ for a simple spread kernel, for example where offspring mutate uniformly within $u/2$ of their parents. To see if this effect could be replicated without the boundary effect, we considered creating a virtual boundary by sharply decreasing the fitness around a shrunk version of the fitness curve. However, initial attempts led to maximum eigenvalues that monotonically decreased with mutation rate. 

\begin{figure}[h]
\centering
 \includegraphics[width=6.5in]{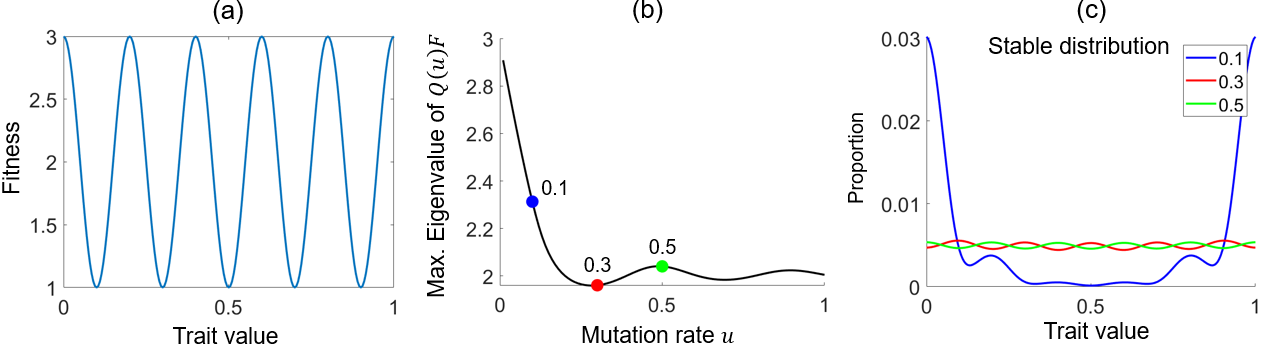}
\caption{A periodic fitness function, (a), can lead to oscillations in the maximum eigenvalue, (b), creating both stable and unstable equilibria. The stable distributions, (c), are mostly periodic, though they peak around the boundary for low mutation rate. }
\label{circle}
\end{figure}

Even when there are no boundary effects, achieved by using a circular trait space, we still found fitness landscapes leading to interesting effects in the evolution of mutation rates, shown in Fig.~\ref{circle} in the same manner as Figs.~\ref{discrete} and \ref{trunc}. Specifically, this occurred for highly periodic fitness functions and the spread kernel where offspring mutate uniformly within an interval of length $u$ around their parent's strategy. Panel (b) shows this setup leads to unstable equilibria, local minima of the maximum eigenvalue plot, but also stable equilibrium not seen in the earlier models. We suspect this occurs because of a wraparound effect, that is, for $u > 1$, the mutants wrap around the boundary, leading to a cluster at the diametrically opposite trait. As $u$ increases further, the offspring once again become more prevalent near the parent's trait. Since one can think of the fitness on the circle as a periodic fitness on the infinite line, the same effect can be achieved for smaller mutation rates by rescaling the fitness function, that is, increasing its frequency. This is significant, as it demonstrates this effect can occur without unnaturally wrapping around the trait space. Further, by increasing the frequency even more, one should be able to create a local maximum in $\lambda_{\max}Q(u)F$ arbitrarily close to $u=0$. This means that stable and unstable equilibria may be reached regardless of initial mutation rate, given an appropriate fitness function. Interestingly, this effect does not occur with the normal spread kernel, as the small tails prevent a similar wraparound phenomenon from occurring. It is also surprising that the stable distribution shown is centered around the boundaries, though some peak is to be expected given the low mutation rate. Lastly, we found that the fitness doesn't need to be strictly periodic, but can become constant near the boundary, and this effect will still occur, albeit to a lesser extent. Not only is this more realistic, but it also leads to a bounded set of traits in the population.



\begin{figure}[h]
\centering
 \includegraphics[width=6.5in]{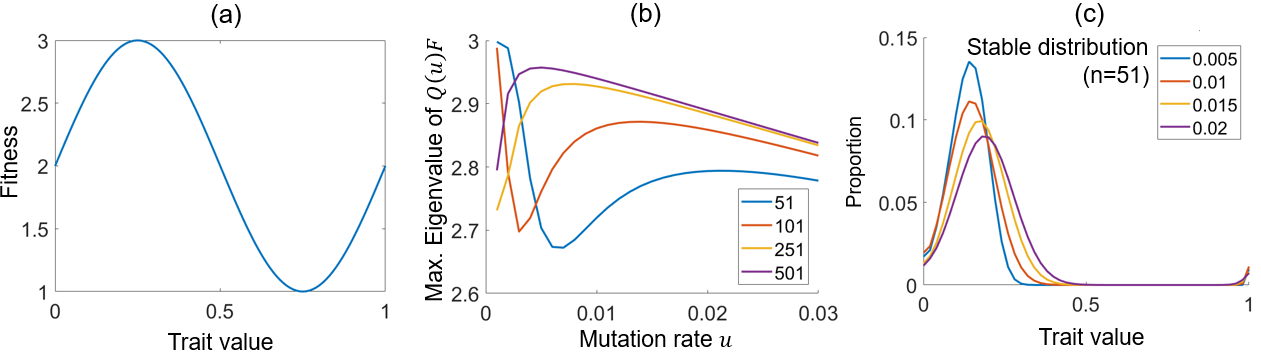}
\caption{A simple fitness function, (a), yields a local optimum near zero. However this moves as the discretization becomes finer, shown as different lines in (b), as opposed to the corresponding plots in the previous figures, where only one discretization is used (since these do not change with the number of traits used). The final panel shows how the stable distributions fail to capture the optimal fitness value, possibly from a bias due to the coarse discretization. }
\label{many_n}
\end{figure}

Lastly, we found local stable and unstable equilibria for low mutation rates using a simple function, using the circular trait space and a normal spread function, depicted in Fig.~\ref{many_n}. One thing to note is that the mutation rate is low enough that only a few entries per row of $Q(u)$ are nonzero, that is, the spread kernel is not well approximated. This is not a problem, as mutants are still accumulated to the closest trait. However this might make the spread kernel effectively biased, which is consistent with the stronger effect seen in coarser discretizations. Interestingly, this effect does not appear in either interval model of mutation. In addition, a different curve is found for a rotated fitness function, which should not be the case, due to the symmetry of the circular model. This may also be an effect of the discretization, as even if the whole fitness is rotated, the particular values may not be, e.g. if the rotation is not a multiple of the spacing. Taken together, these results suggest the importance of specifying the mutation model considered, and an intricate trade-off in the discretization of continuous traits. These issues are promising for further investigation, especially with more complicated fitness functions.


\section{Conclusion}
In this work, we examined how mutation rate would evolve in various circumstances. To do this, we employed adaptive dynamics and the replicator-mutator equation, following the work of \cite{main_ref}. Specifically, we compute the maximum eigenvector of the matrix $Q(u)F$ for various values of $u$ to determine the path of evolution. In several models of mutation, we found fitness functions that lead to nontrivial evolution of mutation rate. 

First we represent traits as finite strings on a finite alphabet. This is inspired by the possibility of multiple alleles of a single gene, so could be a good model of reality, in simple cases of inheritance. Additionally, the closed form of $Q(u)F$  makes the computations simpler, though it is still unfeasible to determine the maximum eigenvalue for arbitrary $F$. Then, we examine local models of mutation, which treats mutation as an incremental change. The challenge in this model is how to most realistically handle the out of bounds mutation, made necessary by the finite nature of the theoretical framework. Accumulation represents a failed attempt to mutate beyond the boundaries, resulting in a trait value at the boundary, whereas truncation effectively ignores such mutations. The issue of mutation beyond the boundaries can be avoided by making the trait space circular, though a circle may be a less natural way of representing traits. In all of these models, we found fitness functions that led to regions where mutation rate will evolve upwards, contrary to intuition. In most cases, this occurred for relatively large mutation rates. 

While this work examined the consequences of a constant fitness function, the method used is easily applicable to other spread kernels and trait topologies. For example, one could consider concave spread kernels, local mutation in higher dimensions, or mutation on a graph, as in \cite{graphs_1}. The same framework can also be applied when fitness varies with frequency or time. For example, by encoding frequency dependent selection using a two player symmetric game played in a well mixed population, we found no selection or selection for lower mutation rates, and observed sensitivity to initial conditions and game parameters. Using a fitness with a single peak that oscillated at various frequencies, we found a sharp transition between selection for low and high mutation, and complicated cycles in the population. In this work, we show that despite simple models, the evolutionary dynamics of mutation rate can not be easily predicted. 

\section*{Acknowledgements}
B.M. is supported by a Dartmouth Fellowship. F.F. is supported by the Bill \& Melinda Gates Foundation (award no. OPP1217336), the NIH COBRE Program (grant no. 1P20GM130454), a Neukom CompX Faculty Grant, the Dartmouth Faculty Startup Fund, and the Walter \& Constance Burke Research Initiation Award. 

\section*{Competeing Interests}
The authors have no competing interests to declare. 

\newpage

\printbibliography 

\end{document}